\documentclass[
a4paper,
aps,
prd,
preprint,
preprintnumbers,
showkeys,
nofootinbib
]{revtex4-1}

\usepackage{braket}
\usepackage{fullpage}
\usepackage{amsfonts}
\usepackage{amsmath}
\usepackage{amssymb} 
\usepackage{graphicx}
\usepackage{epic}
\usepackage{eepic}
\usepackage{epsfig}
\usepackage{latexsym}
\usepackage{color}
\usepackage{float}
\usepackage{multirow} 
\usepackage[citecolor=green]{hyperref} 
\usepackage[caption=false]{subfig}

\usepackage{booktabs}
\usepackage{shorthand}

\begin{document}

\title{Constraining minimal anomaly free $\mathrm{U}(1)$ extensions of the Standard Model}

\author{Andreas Ekstedt}
\email{andreas.ekstedt@physics.uu.se}

\author{Rikard Enberg}
\email{rikard.enberg@physics.uu.se}

\author{Gunnar~Ingelman}
\email{gunnar.ingelman@physics.uu.se}

\author{Johan L\"ofgren}
\email{johan.lofgren@physics.uu.se}

\author{Tanumoy Mandal}
\email{tanumoy.mandal@physics.uu.se}
\affiliation{Department of Physics and Astronomy, Uppsala University, Box 516, SE-751 20 Uppsala, Sweden}

\date{\today}

\begin{abstract} 
We consider a class of minimal anomaly free $\mathrm{U}(1)$ extensions of the Standard Model with three generations of right-handed neutrinos and a complex scalar. 
Using electroweak precision constraints, new 13 TeV LHC data, and considering theoretical limitations such as perturbativity, we show that it is possible to constrain a wide class of models. By classifying these models with a single parameter, $\kp$, we can put a model independent upper bound on the new $\mathrm{U}(1)$ gauge coupling $g_z$. We find that the new dilepton data puts strong bounds on the parameters, especially in the mass region $M_{Z'}\lesssim 3~ \mathrm{TeV}$.
\end{abstract}



\maketitle

\section{Introduction}
\label{sec:intro}
Many extensions of the Standard Model (SM) predict a massive, electrically neutral, color singlet gauge boson (in general called $Z'$) at the TeV scale or higher. Examples include grand unified theories~\cite{Langacker:1980js,Robinett:1981yz,Robinett:1982tq,Langacker:1984dc,London:1986dk,Hewett:1988xc},
string theoretical models~\cite{Ibanez:1987sn,Casas:1987us,Cvetic:1995rj,Cvetic:1996mf}, 
extra-dimensional models~\cite{Masip:1999mk,Casalbuoni:1999ns,Delgado:1999sv,Appelquist:2000nn,Agashe:2003zs,
Agashe:2007ki}, theories of new strong
dynamics~\cite{Hill:1994hp,Hill:2002ap}, little Higgs models~\cite{ArkaniHamed:2001nc,ArkaniHamed:2002qx,Han:2003wu}, and various Stueckelberg extensions~\cite{Kors:2004dx,Kors:2004ri,Kors:2005uz,Feldman:2007wj}. For reviews on $Z'$ phenomenology see~\cite{Leike:1998wr,Langacker:2008yv,Rizzo:2006nw,Basso:2011hn}. For this reason, the ATLAS and the
CMS collaborations have searched for $Z'$ bosons in various channels, including at the 13 TeV LHC~\cite{Khachatryan:2015dcf,ATLAS:2015nsi,CMS:2015nhc,ATLAS:2015nhc,ATLAS:2015dib}. No confirmation or hint of a $Z'$ has been found so far. Nevertheless, an excess at a mass of around 2 TeV in diboson resonance searches by the
ATLAS collaboration~\cite{Aad:2015owa} garnered excitement for some time. 

In many of these experimental searches it is assumed that $Z'$ has a sequential-type ``model independent'' parametrization of its couplings. For example, CMS has obtained a lower limit of 3.15 TeV  on the mass of $Z^{'}$ in the dilepton channel, assuming a sequential $Z'$~\cite{CMS:2015nhc}. A similar mass
limit of 3.4 TeV on a sequential $Z'$ is obtained by ATLAS using 13 TeV dilepton resonance search data~\cite{ATLAS:2015nhc}. 
There are also strong field theoretical requirements such as anomaly cancellation and perturbativity that can severely restrict the parameter space of various
$Z'$ models. 

In this paper we investigate the possible parameter space of a class of minimal $\mathrm{U}(1)$ extensions of the SM that predict a $Z'$ gauge boson, by considering anomaly cancellation, electroweak precision constraints and direct collider limits. The assumptions of our approach are (i) the existence of an additional $\mathrm{U}(1)$ gauge group which is broken by the vacuum expectation value (VEV) of a complex scalar, (ii) the SM fermions are the only fermions that are charged under the SM gauge group, (iii) there are three generations of right-handed neutrinos which are SM singlets but charged under the new $\mathrm{U}(1)$, (iv) the right-handed neutrinos obtain masses via a Type-I seesaw scenario, (v) the gauge charges are generation independent, and (vi) the electroweak symmetry breaking (EWSB) occurs as in the SM. 
The cancellation of the gauge anomalies places a strong theoretical constraint on the theory. If they are not canceled, the theory will not necessarily be unitary or renormalizable, and will have to be considered as an effective theory. 

This paper is organized as follows:\ In Section~\ref{sec:SymmBreak}, we briefly review the
gauge, scalar and fermion sectors of a generic $\mathrm{U}(1)$ extension of the SM; in Section~\ref{sec:charges} we discuss the anomaly cancellation conditions and charge assignments of 
various fields under the new $\mathrm{U}(1)$ gauge group;
in Section~\ref{sec:models} we briefly discuss a few specific $\mathrm{U}(1)$ extended models and introduce a generic
anomaly-free $\mathrm{U}(1)$ model parametrization.
In Section~\ref{sec:DWBR} we present the analytical formulas for various 
decay modes of $Z'$ and show branching ratios (BRs) for some specific models. 
In Section~\ref{sec:results} we discuss the exclusion limits on model parameters from experimental constraints and electroweak precision tests (EWPT). Finally, we present our conclusions in Section~\ref{sec:discussion}.

\section{A brief review of the $\mathbf{U(1)}$ extension}
\label{sec:SymmBreak}

In this section, we review the gauge, scalar and fermion sectors of a generic $\mathrm{U}(1)$ extension of the SM, following mostly the notations and conventions of Ref.~\cite{Appelquist:2002mw}. In general, when a gauge theory consists of several $\mathrm{U}(1)$ gauge groups, kinetic mixing becomes possible. However, this mixing can be rotated away at a given scale. Hence, we can employ a framework where kinetic mixing is not present at tree-level, but which has to be properly taken care of at loop-level.

\emph{A priori} there are two options for the gauge group structure and the subsequent symmetry breaking pattern. One option is to start from the group $\mathrm{SU}(3)_C \times \mathrm{SU}(2)_L \times \mathrm{U}(1)_Y \times \mathrm{U}(1)_z$ and to break the  $\mathrm{U}(1)_z$ group at a high scale while breaking $\mathrm{SU}(3)_C \times \mathrm{SU}(2)_L \times \mathrm{U}(1)_Y$ at the EWSB scale as in the SM. Another option is to consider the gauge group $\mathrm{SU}(3)_C \times \mathrm{SU}(2)_L \times \mathrm{U}(1)_1 \times \mathrm{U}(1)_2$, and to first break $\mathrm{U}(1)_1 \times \mathrm{U}(1)_2$ down to $\mathrm{U}(1)_Y$ at a high scale, and then proceed with the standard EWSB. 
However, it turns out that these possibilities of symmetry breaking are equivalent. It is always possible by redefining the gauge fields and rescaling the gauge couplings to make the $\mathrm{U}(1)_1 \times \mathrm{U}(1)_2$ group look like $\mathrm{U}(1)_Y \times \mathrm{U}(1)_z$ (see Ref.~\cite{Appelquist:2002mw} for a discussion on this point). 

Being equivalent, both symmetry breaking patterns result in the usual SM gauge bosons with an additional electrically and color neutral heavy gauge boson, which we denote as $Z'$.
If the high scale symmetry breaking occurs at the TeV scale we expect the mass of $Z'$ to be at the TeV-scale, and hence it might be observed at the LHC. 
Without any loss of generality we present our model setup by considering the gauge structure $\mathrm{SU}(3)_C \times \mathrm{SU}(2)_L \times \mathrm{U}(1)_Y \times \mathrm{U}(1)_z$ as a template for a minimal $\mathrm{U}(1)$ extension of the SM. 

\subsection{Gauge sector}
\label{ssec:GaugeSector}

We consider the spontaneous symmetry breaking of $\mathrm{U}(1)_z$ by an SM singlet complex scalar field $\varphi$ that acquires a VEV $v_\varphi$.
The charge of this scalar under $\mathrm{U}(1)_z$ can be scaled to $+1$ by redefining the
$\mathrm{U}(1)_z$ coupling $g_z$. The Higgs doublet $\Phi$ responsible for EWSB can in general be charged  under $\mathrm{U}(1)_z$. This leads to a mixing between the $Z$ and $Z'$ bosons after symmetry breaking. With these conventions, the kinetic terms for 
$\Phi$ and $\varphi$ can be written as
\begin{align}\label{eq:CovariantDeriv}
\left|\left(\partial^\mu -i\frac{g}{2}W^\mu -i\frac{g'}{2}B_Y^\mu-i z_H \frac{g_z}{2}B_z^\mu\right)\Phi\right|^2 +\left|\left(\partial^\mu -i\frac{g_z}{2}B_z^\mu\right)\varphi\right|^2 ,
\end{align}
where $z_H$ is the charge of $\Phi$ under $\mathrm{U}(1)_z$. The gauge fields associated with $\mathrm{SU}(2)_L$, $\mathrm{U}(1)_Y$ and $\mathrm{U}(1)_z$ are $W^{\mu},B_Y^{\mu}$ and $B_z^\mu$, with gauge couplings $g,g'$ and $g_z$ respectively. Denoting the VEVs of $\Phi$ and $\varphi$ by $v_H$ and $v_\varphi$ respectively, the relevant mass terms (omitting $W^{\pm}$) after EWSB are
\begin{align}\label{eq:MassMatrix}
\frac{v_H^2}{8}\left(g W^{3\mu}-g' B_Y^\mu -z_H g_z B_z^\mu\right)^2+\frac{v_\varphi^2}{8}g_z^2 B_z^\mu B_{z\mu}\ ,
\end{align}
where $v_H\approx 246~\textrm{GeV}$. 
If $z_H \neq 0$, the diagonalization of the mass matrix will introduce mixing between the SM $Z$ boson and the new $\mathrm{U}(1)_z$ $Z'$ boson, characterized by a mixing angle $\theta'$. Defining $t_z \equiv g_z/g$, $\tan\theta_w \equiv g'/g$ and $r\equiv v_\varphi^2/v_H^2$, the  gauge fields ($B_Y^\mu,~ W^{3 \mu},~B_z^\mu)$ can, for $z_H\neq 0$, be written in terms of the physical fields as
\begin{align}\label{eq:Eigenstates}
\begin{pmatrix}
B_Y^\mu\\
W^{3 \mu}\\
B_z^\mu
\end{pmatrix}
=\begin{pmatrix}
\cos\theta_w & -\sin\theta_w \cos \theta' & \sin\theta_w \sin\theta'\\
\sin\theta_w & \cos\theta_w \cos\theta' & -\cos\theta_w\sin\theta' \\
0 & \sin\theta' & \cos\theta'
\end{pmatrix}
\begin{pmatrix}
A^\mu\\
Z^\mu\\
Z'^\mu
\end{pmatrix},
\end{align}
where $\theta_w$ is the Weinberg angle, and the $Z\leftrightarrow Z'$ mixing angle $\theta'$ is given by 
\begin{align}
\label{eq:mixang}
\theta'=\frac{1}{2}\arcsin\left(\frac{2 z_H t_z c_w}{\sqrt{\lt[2 z_H t_z c_w\rt]^2+\lt[(r+z_H^2)t_z^2c_w^2-1\rt]^2}}\right) .
\end{align}
In the above expression, we use the abbreviation $\cos\theta_w\equiv c_w$.
After symmetry breaking the photon field $A^{\mu}$ remains massless, while the other two physical fields $Z$ and $Z'$ acquire masses which are given by
\begin{align}\label{eq:ZZ'mass}
M_{Z,Z'}=\frac{g v_H}{2 c_w}\left[\frac{1}{2}\lt\{(r+z_H^2)t_z^2c_w^2+1\rt\} \mp \frac{z_H t_zc_w}{\sin2\theta'}\right]^\frac{1}{2} .
\end{align}
In this paper we are interested in the case $M_{Z'}>M_Z$ and from now on we assume this is the case. Due to the induced mixing between $Z$ and $Z'$, the $Z$ couplings are in general different from the SM $Z$-couplings. Therefore, $Z$-couplings measurements can place severe bounds on these models. An observable sensitive to the $Z$-couplings is its width, which is
very precisely measured. 
In Section~\ref{sec:results}, we use the value $\Gamma_Z=2.4952\pm 0.0023$ GeV taken from 
Ref.~\cite{Agashe:2014kda} to constrain the parameter space of the $\mathrm{U(1)}_z$ models.

The gauge sector has, when compared to the SM gauge sector, five new quantities $(g_z, z_H,M_{Z'},\theta',v_\varphi)$. However, Eq.~\eqref{eq:mixang} and the $M_{Z'}$ equation in~\eqref{eq:ZZ'mass} can be used to express two of these parameters in terms of the three remaining free parameters. In principle, it is also possible to use the $M_{Z}$-equation in~\eqref{eq:ZZ'mass} to express a third parameter in terms of $M_Z$ (and other SM parameters) and the two remaining free parameters. However, Eq.~\eqref{eq:ZZ'mass} 
is a tree-level relation and the measured $M_Z$ is slightly different from its SM tree-level prediction. This difference is due to higher-order effects and new physics, if it is present.
We observe that expressing $z_H$ (or the product $g_z z_H$) by using the (tree-level) $M_Z$-equation
in \eqref{eq:ZZ'mass} makes $z_H$ very sensitive to this difference. Therefore, we cannot
use the tree-level $M_Z$-equation to reduce the number of free parameters from three to two. 
Instead one should really use the BSM mass relation of $M_Z$ in Eq.~\eqref{eq:ZZ'mass} which induces a tree level contribution to the oblique parameters. In particular, the tree level contribution to the $T$-parameter is~\cite{Appelquist:2002mw}
\begin{align}
\alpha T=\frac{\Pi^{new}_{ZZ}}{M_{Z}^{2}}=\frac{M_{Z}^{2}-(M_{Z}^0)^{2}}{M_{Z}^{2}}\ ,
\end{align}
where $M_{Z}$ is the prediction of the $Z$ mass from equation~\eqref{eq:ZZ'mass}, $M_{Z}^0=g v_{H}/(2 c_{w})$ is the corresponding SM tree-level prediction, and $\alpha$ is the fine-structure constant evaluated at the $Z$-pole.
There will be additional loop corrections to the $T$-parameter, but these are suppressed by the mixing angle and can be neglected. 
The current measured value of the $T$-parameter is $0.05\pm 0.07$ \cite{Agashe:2014kda}
and we use this value in our analysis.

In the end, we have three free parameters, which we take to be $\{z_H, g_z,M_{Z'}\}$. However,
in the observables we consider in our analysis, $z_H$ and $g_z$ always show up as a product.
Therefore, one can effectively consider $\{z_H g_z,M_{Z'}\}$ as the set of free parameters in this model. 
We define
$\mc{A}(M_{Z'})\equiv 8 c_w^2 M_{Z'}^2/(g^2 v_H^2)$
for convenience, and find an expression for $v_\varphi$ in terms of $\{z_H, g_z,M_{Z'}\}$ from Eqs.~\eqref{eq:mixang} and~\eqref{eq:ZZ'mass}, 
\begin{align}
v_\varphi^2&=v_H^2 \mc{A}(M_{Z'}) \frac{ \left\{\mc{A}(M_{Z'})-2-2 c_w^2 t_z^2 z_H^2\right\}}{2c_w^2 t_z^2\left\{\mc{A}(M_{Z'})-2\right\}}\equiv v_\varphi^2(z_H, g_z,M_{Z'})\ .
\label{eq:phiVEV}
\end{align}
We can then employ the parametrization of Eq.~\eqref{eq:phiVEV} together with Eq.~\eqref{eq:mixang} to express the mixing angle $\theta'$ as a function of $M_{Z'},z_H$ and $ g_z$; similarly we express $M_Z$ in terms of these parameters.
Using this parametrization we place restrictions on the parameter space using collider data, $T$ parameter constraints and $\Gamma_Z$ constraints in Section~\ref{sec:results}.

\subsection{Scalar sector}

The new complex scalar field $\varphi$, introduced in order to break the 
$\mathrm{U(1)}_z$ symmetry, leads to the possibility of a more general scalar potential. The most general gauge invariant and renormalizable potential can be written in the form 
\begin{align}
\label{eq:ScalarPot}
V= -\mu^2_{\Phi} \lt(\Phi^\dag\Phi\rt) - \mu^2_{\varphi} \lt|\varphi\rt|^2 + \lambda_1 \lt(\Phi^\dag\Phi\rt)^2 + \lambda_2  \left(\lt|\varphi\rt|^2\right)^2 +\lambda_3 \lt(\Phi^\dag\Phi\rt) \lt|\varphi\rt|^2 .
\end{align}
This potential has 5 free parameters. For this potential to be responsible for the symmetry breaking, it has to be bounded from below, and it must have a global minimum located away from the origin. To be bounded from below, the parameters of the potential have to satisfy the following two conditions~\cite{Basso:2011hn}
\begin{align}
\label{eq:BoundedPot}
\lambda_1, \lambda_2 > 0;\qquad 4 \lambda_1 \lambda_2 - \lambda_3^2 > 0\ .
\end{align}
For the purpose of minimization it is convenient to work in the unitary gauge, in which the VEVs of the scalar fields can be written as
\begin{align}
\label{eq:UnitaryGauge}
\braket{\Phi} \equiv \frac{1}{\sqrt{2}}\binom{0}{v_\text{H}};\qquad
\braket{\varphi} \equiv \frac{v_\varphi}{\sqrt{2}} \ .
\end{align}
By requiring the potential to be minimized away from the origin, for the fields $\Phi$ and $\phi$ to acquire their VEVs, the parameters $\mu^2_{\Phi},\mu^2_{\varphi}$ in the potential can be expressed in terms of the VEVs, by the following relations
\begin{align}
\label{eq:MassEig}
\mu^{2}_{\Phi}=2\lambda_1 v_\Phi^{2}+\lambda_3 v_\varphi^{2};\qquad\mu^{2}_{\varphi}=2\lambda_2 v_\varphi^{2}+\lambda_3 v_\Phi^{2}.
\end{align}

Note that the introduction of a new complex scalar field will in general result in mixing between the SM Higgs boson and the new scalar state. The five parameters introduced in Eq.~\eqref{eq:ScalarPot} can then be expressed in terms of the VEVs $v_{H}$ and $v_{\varphi}$, the masses of the physical scalars $M_{H_1}$ and $M_{H_2}$, and the sine of the mixing angle between $H_1$ and $H_2$ denoted by $\sin\alpha$. Using Eq.~\eqref{eq:MassEig}, we obtain the following relations
\begin{align}\label{eq:ScalarPhysParam}
\lm_1 = \frac{M_{H_1}^2c_{\al}^2 + M_{H_2}^2s_{\al}^2}{2v_H^2};\qquad
\lm_2 = \frac{M_{H_1}^2s_{\al}^2 + M_{H_2}^2c_{\al}^2}{2v_{\varphi}^2};\qquad
\lm_3 = \frac{\lt(M_{H_2}^2 - M_{H_1}^2\rt)s_{\al}c_{\alpha}}{v_Hv_{\varphi}} \ ,
\end{align}
where we use the shorthand notations $s_\al \equiv \sin\alpha;~c_\al \equiv \cos\alpha$ and we follow the conventions $M_{H_2}\geq M_{H_1}$ and $-\pi /2 \leq \alpha \leq \pi /2$.
We take $v_H = 246$~GeV and $M_{H_{1}}=125$ GeV.\footnote{By choosing instead $M_{H_2}=125$ GeV, one can consider the possibility that there is a lighter scalar yet to be found at the LHC. We will not be concerned with this since we do not study the scalar sector in detail.} Then in the scalar sector we only have two free parameters that are not determined from the SM or the gauge sector, which we choose to be $M_{H_2}$ and $\sin\alpha$. Note that for a given $M_{Z'}$, $v_\varphi$ is given as a function of $g_z$ and $z_H$.

\subsection{Fermion sector}
Apart from the SM fermions we also introduce three generations of right-handed neutrinos, required to cancel various gauge anomalies which we discuss in the following subsection. The three generations of left-handed quark and lepton doublets are denoted by $q_L^i$ and $l_L^i$ respectively and the right-handed components of up-type, down-type quarks and
charged leptons are denoted by $u_R^i$, $d_R^i$ and $e_R^i$ (here $i=1,2,3$) respectively; the three right-handed neutrinos are denoted as $\nu_{R}^k$. All the SM fermions are, in general, charged under the $\mathrm{U}(1)_z$ group and the right-handed neutrinos are singlets under the SM gauge group but charged under $\mathrm{U}(1)_z$. The $\mathrm{U}(1)_z$ charges are determined from the Yukawa couplings and the anomaly cancellation conditions, which require that the right-handed neutrinos are charged under $\mathrm{U(1)}_z$. The anomaly cancellation conditions will be elaborated in the following section.

For definiteness we assume that neutrino masses arise from the
type-I seesaw scenario, by allowing Majorana mass terms to be generated from the $\mathrm{U(1)}_z$ breaking. Dirac mass terms are then generated from EWSB, and upon diagonalization we obtain 3 light and 3 heavy Majorana states. We restrict ourselves to the case of small mixing between generations, since this will not affect $Z'$ phenomenology. This mixing would be important for a dedicated study of the neutrino sector, but this is
beyond the scope of the present paper.

In principle, mixing between the left and right-handed neutrinos could be important. For type-I seesaw the mixing angle is given by 
\begin{align}
\frac{1}{2} \arctan\left[-2\frac{\sqrt{M_{\nu_R} M_{\nu_L}}}{M_{\nu_R} +M_{\nu_L}}\right]\sim -\sqrt{\frac{M_{\nu_L}}{M_{\nu_R}}}\ ,
\end{align}
where $M_{\nu_L}$ and $M_{\nu_R}$ are the masses of the left-handed and right-handed neutrinos respectively.
Since the left-handed neutrinos have extremely small masses, this mixing is not important for the $Z'$ phenomenology considered in this paper.

\section{Anomaly cancellation \& $\mathbf{U(1)_z}$ charges}
\label{sec:charges}

We wish to consider here a class of anomaly free models and what restrictions anomaly cancellation places on the spectrum of possible fermion charges. 

To construct an anomaly-free gauge theory with chiral fermions, we should assign the gauge charges of the fermions respecting all types of gauge-anomaly cancellation conditions. These conditions arise when contributions from all anomalous triangle diagrams are required to sum to zero. There are six types of possible anomalies as listed below, leading to six conditions that have to be satisfied in order to make the theory anomaly-free:
\bi
\item The $\lt[\mathrm{SU}(2)_L\rt]^2\lt[\mathrm{U}(1)_z\rt]$ anomaly, which implies $\textrm{Tr}\lt[\lt\{T^i,T^j\rt\}z\rt]=0$,

\item The $\lt[\mathrm{SU}(3)_c\rt]^2\lt[\mathrm{U}(1)_z\rt]$ anomaly, which implies $\textrm{Tr}\lt[\lt\{\mc{T}^a,\mc{T}^b\rt\}z\rt]=0$,

\item The $\lt[\mathrm{U}(1)_Y\rt]^2\lt[\mathrm{U}(1)_z\rt]$ anomaly, which implies $\textrm{Tr}\lt[Y^2z\rt]=0$,

\item The $\lt[\mathrm{U}(1)_Y\rt]\lt[\mathrm{U}(1)_z\rt]^2$ anomaly, which implies $\textrm{Tr}\lt[Yz^2\rt]=0$,

\item The $\lt[\mathrm{U}(1)_z\rt]^3$ anomaly, which implies $\textrm{Tr}\lt[z^3\rt]=0$,

\item The gauge-gravity anomaly, which implies $\textrm{Tr}\lt[z\rt]=0$.
\ei
The traces run over all fermions. The generators of $\mathrm{SU}(2)_L$ and $\mathrm{SU}(3)_c$ are represented
by $T^i$ and $\mc{T}^a$ respectively, and we denote hypercharge by $Y$ and the $\mathrm{U}(1)_z$ charge by $z$. We assume the charges $z$ to be generation independent, just as for the charges in the SM. Generation dependent charges are in principle not forbidden, but they may lead to flavor changing neutral currents. 
The charges of the fermions are labeled as: $z_{q}$ -- left-handed quark doublets, $z_{u}$ -- right-handed up-type quarks, $z_{d}$ -- right-handed down-type quarks, $z_{l}$ -- left-handed lepton doublets, 
$z_{e}$ -- right-handed charged leptons and $z_{k}$ -- right-handed neutrinos.

By requiring that the EWSB gives mass to the SM fermions, the relations $z_{H}=z_{u}-z_{q}=z_{e}-z_{l}=z_{q}-z_{d}$ must hold for the Yukawa interactions to be gauge invariant~\cite{Appelquist:2002mw}. It should be noted that the mixed gauge anomaly $\lt[\mathrm{SU}(3)_c\rt]^2\lt[\mathrm{U}(1)_z\rt]$ cancels automatically from the Yukawa coupling constraints above.

By requiring that all the other gauge anomalies vanish one can obtain relations between these charges. One finds that~\cite{Appelquist:2002mw}
\begin{align}
& z_{l} = -3z_{q};~~ z_{d} = 2z_{q}-z_{u};~~ z_{e} = -2z_{q}-z_{u}; \label{eq:anferm}\\
& \frac{1}{3}\sum_{k=1}^{n}z_{k} = -4z_{q}+z_{u};~~
\left(\sum_{k=1}^{n}z_{k}\right)^{3} = 9\sum_{k=1}^{n}z_{k}^{3} \label{eq:anoRHN}\ .
\end{align}
It is well known that the most general solution to the anomaly cancellation conditions (in the framework with no kinetic mixing) is for the charge $Q_{f}$ of a given fermion $f$ to be written as a linear combination of its hypercharge $Y_{f}$ and $(B-L)_{f}$ quantum number~\cite{Weinberg2}, i.e., $Q_{f}=aY_{f}+b(B-L)_{f}$. In our convention, this becomes~\cite{Appelquist:2002mw}
\begin{equation}
Q_{f}=(z_{u}-z_{q})Y_{f}+(4z_{q}-z_{u})(B-L)_{f},\label{eq:fermcharge}
\end{equation}
which is consistent with Eq.~\eqref{eq:anferm}. In Table~\ref{tab:charges}, we summarize the gauge charges of all the relevant fields.

\begin{table}
\centering
\vspace{0.5em}
\begin{tabular}{|ccccc|}
\hline
&$\mathrm{SU}(3)_{c}$&$\mathrm{SU}(2)_{L}$&$\mathrm{U}(1)_{Y}$&$\mathrm{U}(1)_{z}$\\
\hline
$q_{L}$&3&2&$1/3$&$z_{q}$\\
$u_{R}$&3&1&$4/3$&$z_{u}$\\
$d_{R}$&3&1&$-2/3$&$2z_{q}-z_{u}$\\
$l_{L}$&1&2&$-1$&$-3z_{q}$\\
$e_{R}$&1&1&$-2$&$-2z_{q}-z_{u}$\\
$\nu_{R}$&1&1&0&$z_{k}$\\
$H$&1&2&$+1$&$-z_{q}+z_{u}$\\
$\varphi$&1&1&0&1\\
\hline
\end{tabular}
\caption{The charge assignments for the fermions and scalars of the model.}\label{tab:charges}
\end{table}

In this model it is possible to introduce Majorana mass terms for the right-handed neutrinos, such as 
$(\varphi^{\dagger})\bar{\nu^{c}}_{R}^{k}\nu_{R}^{k}$,
provided that $z_{k}=1/2$, since $\varphi$ has unit $\mathrm{U(1)}_z$ gauge charge (a mass term is also possible for $z_{k}= -1/2$, but we ignore this since this choice will not provide any different conclusion than the $z_{k}=1/2$ case).  Hence, if we want to be able to have both Majorana and Dirac mass terms from renormalizable interactions 
(i.e., a seesaw mechanism), we require that all the right-handed neutrino charges are equal to $1/2$; from Eq.~\eqref{eq:anoRHN} we then find
\begin{equation}
z_k=4z_q - z_u = 1/2\label{eq:chargeRHN}.
\end{equation}

Including three right-handed neutrinos introduces three new parameters, i.e., the masses of the three right-handed neutrinos. We find that the only influence of the neutrino masses on the phenomenology is whether or not the decay channel $Z' \rightarrow \nu_{R}\nu_{R}$ is open. In this paper, we take the masses to be degenerate and equal to $500$ GeV. This somewhat arbitrarily chosen mass ensures that the $Z' \rightarrow \nu_{R}\nu_{R}$ channel remains open for the entire mass region of interest, while at the same time not being light enough to conflict with experimental neutrino constraints.

In our setup we are only considering the SM fermion content together with right-handed neutrinos.

\section{$\mathbf{U(1)_z}$ models}
\label{sec:models}

So far we have discussed a very wide class of anomaly-free 
$\mathrm{U}(1)$ extensions. Many cases of these models have been studied in the literature~\cite{Staub:2016dxq}. We briefly review some of them here and then introduce a model
independent parametrization for this class of models.

\subsection{Specific models}
\label{ssec:specmodels}

\subsubsection{Gauged $B-L$}
A particularly attractive $\mathrm{U}(1)$ extension of the SM is where the $B-L$ quantum number is gauged, usually called  $\mathrm{U(1)}_{B-L}$. Specifically all fermion charges are proportional to their $B-L$ quantum numbers. From Eq.~\eqref{eq:fermcharge} we see that this corresponds to the choice $z_{u}=z_{q}$. This model can also be thought of as the special case of no $Z\leftrightarrow Z'$ mixing, since $z_{H}=z_{u}-z_{q}=0$, which is the only charge assignment that results in vector-like couplings for the fermions. 
There exist extensive dedicated studies~\cite{Basso:2011hn,Coutinho:2011xb} of the $B-L$ model in the literature to which we refer the reader for a deeper discussion.

\subsubsection{$Y$-sequential}
Another natural model is one where the new gauge charges obey the same relations as the hypercharges. From Eq.~\eqref{eq:fermcharge} we see that this model, known as the Y-sequential model, is achieved when $z_u=4z_q$. An interesting and special feature of this model is that right-handed neutrinos are redundant, since as can be seen from Eq.~\eqref{eq:anoRHN}, anomaly cancellation is ensured without any right-handed neutrinos. In this paper we consider the minimal Y-sequential model, by assuming that there exist no right-handed neutrinos charged under the gauge groups.
It is important to note that the $Y$-sequential model is different from the
so-called sequential Standard Model (SSM), which is not anomaly free.

\subsubsection{$SO(10)$-GUT}
The $\mathrm{SO(10)}$ model is a widely studied model as a candidate of
grand unified theories (GUTs), with and without supersymmetry. One of the possible breaking patterns for the $\mathrm{SO(10)}$ group is to break down to a flipped $\mathrm{SU(5)}$ model, i.e., $\mathrm{SO(10)} \rightarrow \mathrm{SU(5)} \times \mathrm{U(1)}$. These models can then upon breaking at a high scale result in a $\mathrm{U(1)}$ extension  surviving after the $\mathrm{SU(5)}$ breaking. These models commonly include new exotic fermionic states, but for the purpose of studying the minimal $\mathrm{U(1)}$ extension these states will be ignored. The model is often denoted as 
$\mathrm{U(1)}_\chi$ and in our framework it is distinguished by the relation $z_q = -z_u$.

\subsubsection{Right-handed}
In the right-handed model, the gauge field corresponding to the new $\mathrm{U(1)}_R$ only couples to the right-handed fermion fields. The gauge charges are proportional to the eigenvalues of the approximate global $\mathrm{SU}(2)_{R}$ symmetry of the SM. This corresponds to the case when $z_q =0$.

\subsubsection{Left-right model}
A neutral heavy gauge boson $Z'$ can originate from left-right symmetric models with gauge group  
$\mathrm{SU}(2)_L\otimes \mathrm{SU}(2)_R\otimes \mathrm{U}(1)_{B-L}$. In addition to $Z'$, 
there are also two massive charged gauge bosons $W'^{\pm}_R$. By redefining
gauge couplings and fields we can always write
$\mathrm{U}(1)_R\otimes \mathrm{U}(1)_{B-L} \equiv \mathrm{U}(1)_Y\otimes \mathrm{U}(1)_z$.
In terms of the $z_H, g_z$ and the fermion charges, this model can be defined by the relations
\be 
z_q = -\frac{g_Y^2}{3g_z^2z_H};~~~z_u = z_H - \frac{g_Y^2}{3g_z^2z_H}.
\ee

\subsection{$\kappa$-parametrization}

All of these particular cases of the $\mathrm{U(1)}$ extensions we discuss above have one thing in common: the charges $z_u$ and $z_q$ can be written as
\begin{equation}
z_q = \kp z_u,
\end{equation}
where $\kp$ is a parameter we introduce in order to present our results in a model-independent fashion. In Table~\ref{tab:cases} we use the $\kappa$-parametrization to summarize some of the specific models considered in subsection~\ref{ssec:specmodels}. We have not included the left-right model since it is not easy to express in this framework; an ambiguity arises since there exist two branches of 
$\kappa$ as functions of $g_z$. In the limit $g_Z\rightarrow \infty$, one branch approaches the right-handed model, and the other one approaches the $B-L$ model. We will hence not study this model separately and instead focus separately on the limiting cases. The charges $z_q$ and $z_u$ can, by their relation to the charge $z_{H}$, be written as
\be
z_q = \frac{\kp}{1-\kp}z_H;~~ z_u = \frac{1}{1-\kp}z_H \ . \label{eq:zqzukappa}
\ee
\begin{table}[H]
\centering
\vspace{0.5em}
\begin{tabular}{|cc|}
\hline
Model & $\kp=z_q/z_u$\\
\hline
Gauged $B - L$ & 1 \\
Y sequential & 1/4 \\
$\mathrm{SO}(10)$-GUT & $-1$\\
Right-handed & 0 \\
\hline
\end{tabular}
\caption{The ratio of the charges $z_q/z_u$, i.e., $\kp$ for specific models with an extra $\mathrm{U}(1)$ symmetry.}\label{tab:cases}
\end{table}

Using equation~\eqref{eq:chargeRHN}, i.e.~requiring that we can write a Majorana mass term for the right handed neutrinos, together with Eqs.~\eqref{eq:zqzukappa}, we find that
\be 
z_H(\kp) = \frac{1-\kp}{2(1-4\kp)}~\Rightarrow ~z_u(\kp)=\frac{\kp}{2(1-4\kp)};~~z_q(\kp)=\frac{1}{2(1-4\kp)}\ .
\ee
Note that this parametrization is only allowed if $\kp\neq 1/4$, which reflects the fact that right-handed neutrinos are not necessarily included in the $Y$-sequential model. This case, therefore, has to be treated separately and we perform a separate analysis for this model.

In the $\kappa$-formalism one can parametrize the production
cross section of $Z'$ at the LHC, i.e., $\sg(pp\to Z')$, in terms of $\kp$ as follows,
\begin{align}\label{eq:cspar}
\sg\lt(M_{Z'},g_z,\kp\rt) &= g_z^2 \bigg\{ a_L^u\lt(M_{Z'}\rt)\lt(\frac{\kp}{1-4\kp}\rt)^2 +
a_R^u\lt(M_{Z'}\rt)\lt(\frac{1}{1-4\kp}\rt)^2 \nn \\
&+ a_L^d\lt(M_{Z'}\rt)\lt(\frac{\kp}{1-4\kp}\rt)^2 +
a_R^d\lt(M_{Z'}\rt)\lt(\frac{2\kp -1}{1-4\kp}\rt)^2\bigg\} \ ,
\end{align} 
where $a_{L/R}^u$ ($a_{L/R}^d$) are the contributions from the left/right components of all the up (down) type quarks in the proton. Using this parametrization we can in a compact manner study a wide class of anomaly free $\mathrm{U(1)}$ extensions. 

Generally the most stringent constraints on $Z'$ models come from dilepton events; thus it is worthwhile to study which $\kp$ value minimizes the contributions to this channel, since this will put a model independent constraint on the parameter space $(g_z,~M_{Z'})$. Performing this minimization numerically we find that the minimum of $\sigma(M_{Z'},\kp) \times \Gamma_\text{ll}$ occurs for $0>\kp>-1/2$, with a slight $M_{Z'}$ dependence coming from the relative strength of the different quark contributions to the production cross section. This $\kp_\text{min}$ value then serves as an important benchmark, since if the model is ruled out by dilepton data for given $(g_z,~M_{Z'})$, then all $\kp$ models are automatically ruled out for this parameter point.

\section{Decay widths \& branching ratios}
\label{sec:DWBR}

The $Z'$ has the following two-body decays: $\bar{f}f$ (where $f$ denotes any Dirac fermion), $\nu\nu$ (where $\nu$ denotes any Majorana fermion), $W^+W^-$ and $ZS$ (where $S$ represents any scalar, i.e., $H_1$ or $H_2$). In this paper we only consider the lowest order results from perturbation theory. The tree-level decay widths can be found from each corresponding interaction Lagrangian.

\begin{itemize}

\item \underline{$Z'\to \bar{f}f$ decay mode:} From the interactions
\begin{equation}
\mathcal{L}_{Z'ff} \supset \lt(i\lambda_{L}~\bar{f}_{L}\gamma^{\mu}f_{L}+
i\lambda_{R}~\bar{f}_{R}\gamma^{\mu}f_{R}\rt) Z'_{\mu}\ ,  
\end{equation}
the partial decay width for $Z'\to \bar{f}f$ decay is given by
\begin{equation}
\Gm\lt(Z'\to \bar{f}f\rt)= \frac{N_cM_{Z'}}{24\pi}\sqrt{1-\frac{4M_f^2}{M_{Z'}^2}}
\lt\{\lt(\lm_L^2+\lm_R^2\rt)\lt(1-\frac{M_f^2}{M_{Z'}^2}\rt)+6\lm_L\lm_R\frac{M_f^2}{M_{Z'}^2}\rt\},
\end{equation}
where $\lm_L$ and $\lm_R$ denote the couplings to the left and right handed fermions respectively, $M_f$ is the fermion mass, and $N_c$ is the number of colors of the fermion.

\item \underline{$Z'\to \nu\nu$ decay mode:} From the interaction
\begin{equation}
\mathcal{L}_{Z'\nu_{x}\nu_{x}} \supset i\lambda_x~(\nu^{c})^{T}\gamma^{\mu}P_{x}\nu  Z'_{\mu}\ ,  
\end{equation}
the partial decay width for $Z'\to \nu\nu$ decay is given by
\begin{align}
\Gm\lt(Z'\to \nu_x\nu_x\rt)= \frac{M_{Z'}}{48\pi}\lm_x^2\lt(1-\frac{4M_{\nu_x}^2}{M_{Z'}^2}\rt)^{3/2},
\end{align}
where $\lm_x$ is the coupling to the $x$ chirality and $P_x$ is the corresponding projection operation. The mass of the Majorana fermion is denoted by $M_{\nu_x}$.

\item \underline{$Z'\to W^+W^-$ decay mode:} The $Z'W^+W^-$ coupling arises from the mixing of the gauge fields $W_3^\mu, B^\mu $ and $B_z^\mu$. From the triple gauge boson interaction
\begin{equation}
\label{eq:tripgau}
\mathcal{L}_{Z'W^{+}W^{-}} \supset \lm_{W}Z'_{\mu}(p_1)W^{+}_{\nu}(p_2)W^{-}_{\rho}(p_3),\
\end{equation}
the partial decay width for $Z'\to W^{+}W^{-}$ decay is given by
\begin{align}
\Gm\lt(Z'\to W^+W^-\rt) = \frac{M_{Z'}^5}{192\pi M_W^4}\lm_{W}^2\lt(1-\frac{4M_W^2}{M_{Z'}^2}\rt)^{\frac{3}{2}} \lt( 1 + \frac{20M_W^2}{M_{Z'}^2} + \frac{12 M_W^4}{M_{Z'}^4}\rt),
\end{align}
where $\lambda_{W}$ is the $Z'W^{+}W^{-}$ coupling. The momentum associated with
each gauge field is shown within bracket in Eq.~\eqref{eq:tripgau} and all momenta
 point towards the three-point vertex.

\item \underline{$Z'\to ZS$ decay mode:} From the interaction
\begin{equation}
\mathcal{L}_{Z'ZS} \supset \mu_S~Z'_{\mu}Z^{\mu}S\ ,  
\end{equation}
the partial decay width for $Z'\to ZS$ decay is given by
\begin{align}
\Gm\lt(Z'\to Z S\rt) &= \frac{\mu_S^2 M_{Z'}}{192\pi M_Z^2}
\lt(1 - \frac{\lt(2M_S^2-10M_Z^2\rt)}{M_{Z'}^2}
+ \frac{\lt(M_S^2 - M_Z^2\rt)^2}{M_{Z'}^4}\rt) \nn\\
&\times \lt(1 - \frac{2\lt(M_S^2+M_Z^2\rt)}{M_{Z'}^2}
+ \frac{\lt(M_S^2 - M_Z^2\rt)^2}{M_{Z'}^4}\rt)^{1/2},
\end{align}
where $\mu_S$ is a dimensionful (mass dimension one) cubic coupling.

\end{itemize}

\begin{figure}[H]
\centering
\captionsetup[subfigure]{}
\subfloat[][]{\includegraphics[width=0.42\textwidth]{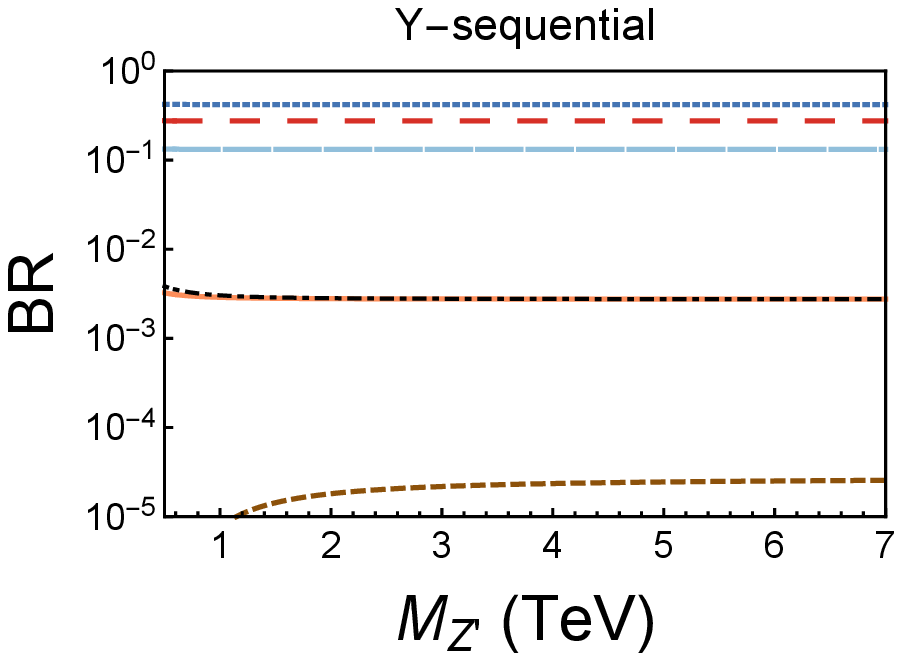}\label{sfig:BRYseq}}~~
\subfloat[][]{\includegraphics[width=0.42\textwidth]{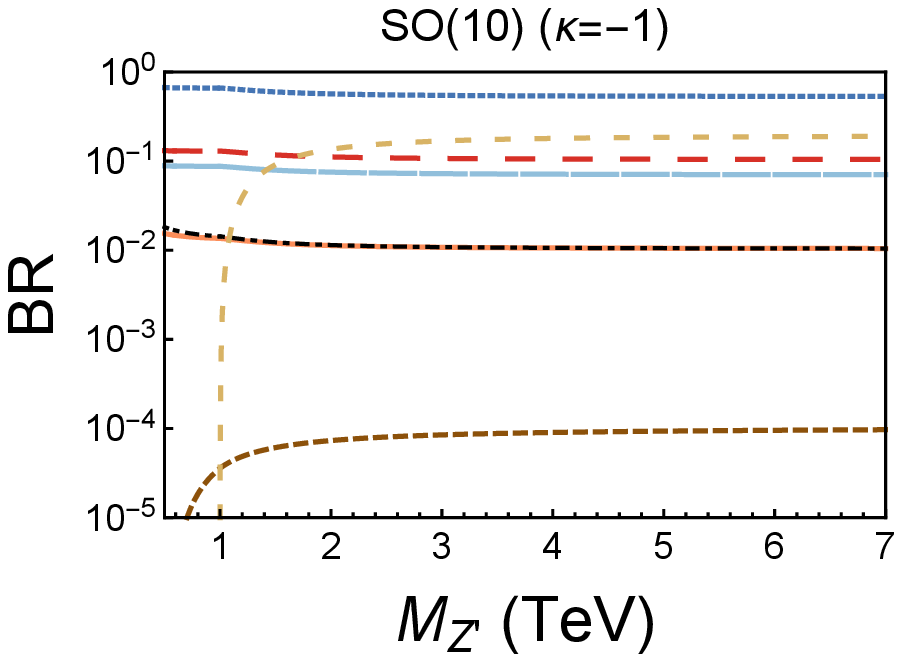}\label{sfig:BRSO10}}
\\
\subfloat[][]{\includegraphics[width=0.42\textwidth]{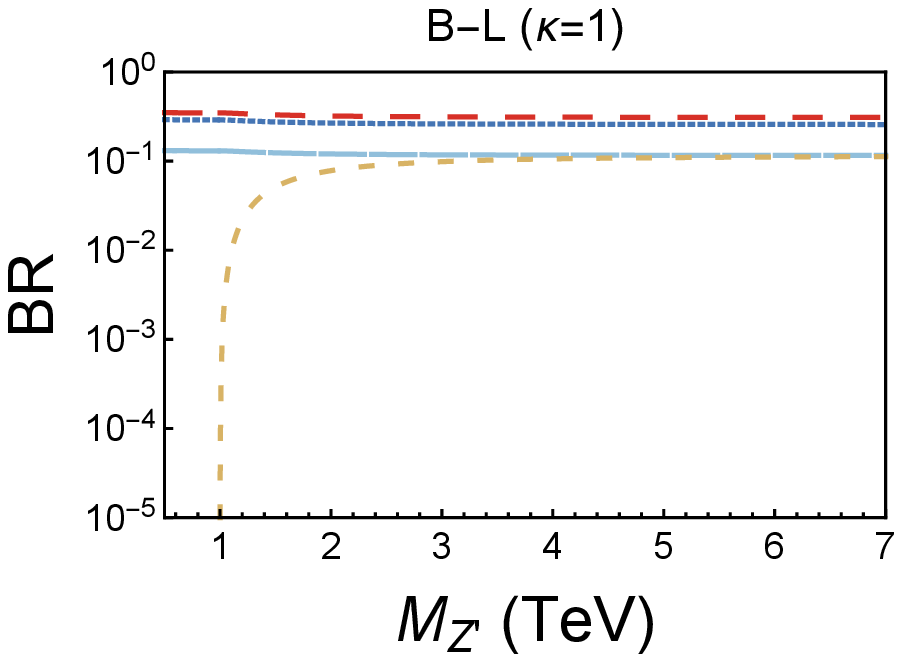}\label{sfig:BRLR}}~~
\subfloat[][]{\includegraphics[width=0.42\textwidth]{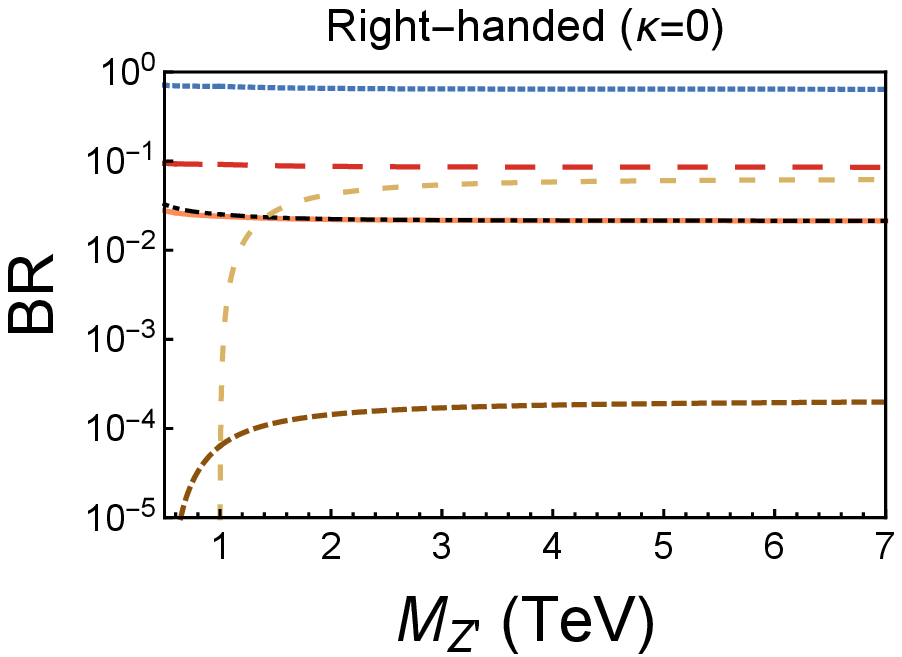}\label{sfig:BRR}}
\\

\hbox{\hspace{27ex}\includegraphics[width=0.5\textwidth]{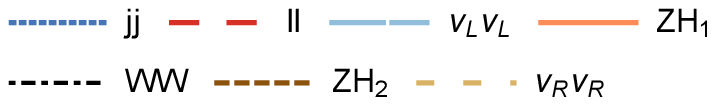}}
\caption{Branching ratios of $Z'$ as functions of $M_{Z'}$ for the models:~(a) 
$Y$-sequential with no right-handed neutrinos, (b) $\mathrm{SO(10)}$-GUT, (c) B-L (d) right-handed model. We use $M_{\nu_R}=500$ GeV, $M_{H_2}=500$
GeV and $\sin\al=0.1$ for all branchings.}
\label{fig:branchings}
\end{figure} 

In Fig.~\ref{fig:branchings} we show the BRs of $Z'$ as functions of the mass
for the specific models we discussed in section~\ref{sec:models}. The final states with biggest BRs are dijets and dileptons. Therefore, in the next section, for the exclusion from experiments we mostly use dilepton and dijet data, where it turns out that the dilepton data is more constraining. We observe that all the BR curves are almost horizontal (after a mode becomes 
kinematically allowed) in the entire range of $M_{Z'}$ in consideration. 
This is because all the couplings of $Z'$ are either constant or depend very weakly on $M_{Z'}$,
 and therefore BRs become insensitive to $M_{Z'}$ since phase-space factors go to a constant value in the $M_{Z'}\to\infty$ limit.
For the right-handed model, the $Z'\to \nu_L\nu_L$ mode is absent since only right-handed fermions couple to $Z'$ in this model. In the $B-L$ model there is no tree-level mixing between $Z$ and $Z'$, and thus there are no direct diboson couplings to $Z'$ at tree level. Notice that $\Gm\lt(Z'\to W^+W^-\rt)\approx \Gm\lt(Z'\to Z H_1\rt)$
for all models, which is a consequence of Goldstone boson equivalence in the 
high energy limit.

\section{Constraints from data}
\label{sec:results}

Using the $\kp$-parametrization described in Section \ref{sec:charges}, 
we perform tree-level calculations of $Z'$ production cross section $\sigma(pp\to Z')$ 
at the LHC using CTEQ6L1~\cite{Pumplin:2002vw} parton distribution
functions (at $\mu_F=\mu_R=M_{Z'}$ where $\mu_F$ and $\mu_R$ are the
factorization and renormalization scales, respectively). 
Various BRs of $Z'$ are calculated analytically using the formulas given in 
section~\ref{sec:DWBR}, where the relevant couplings are obtained using \textsc{FeynRules-2.3}~\cite{Alloul:2013bka}.
The production cross sections are computed 
using~\textsc{MadGraph5\underline{\space}aMC@NLO}~\cite{Alwall:2014hca}. Using the parametrization shown in Eq.~\eqref{eq:cspar}, the (fitted) functional forms of $a^{u,d}_{L,R}(M_{Z'})$ are obtained by interpolating the cross sections. The narrow width approximation is then used in order to write $\sigma\left(pp\rightarrow Z'\rightarrow XY\right)\approx \sigma\left(pp\rightarrow Z'\right)\times BR(Z'\to XY)$.

In this section we will see that the main constraint on minimal $\mathrm{U}(1)$ extensions of the SM comes from the dilepton channel. Since some free parameters of the model ($M_{\nu_R},M_{H_2}$ and $\sin\alpha$) have very little effect on the dilepton branching, we fix them at reasonable values. The only real effect of the mass parameters on the $Z'$ phenomenology is whether the corresponding decay channel is open or not. We choose $M_{\nu_R}=500$ GeV and $M_{H_2}=500$ GeV such that these channels are open in the mass range we study, and $\sin\alpha = 0.1$ motivated by the SM-likeness of the observed Higgs boson.

In order to place exclusion bounds on the models, we compare the 95\%
confidence level (CL) upper limits (UL) on cross sections (the quantity used
is $\sg\times BR$ where $\sg$ is the production cross section and BR denotes the 
branching of $Z'$ in the corresponding channel) 
using dijet and dilepton data from the 13 TeV LHC~\cite{ATLAS:2015nsi,ATLAS:2016cyf}. In our analysis, we use only the ATLAS data since the CMS data puts very similar bounds on the parameter space.
In addition, the models are constrained by EWPT constraints, in particular by tree-level contributions to the $T$-parameter and to the $Z$ width. In principle there is also a bound on $z_H g_Z$ from perturbativity, but this is much less constraining than the bounds from data.

While comparing with the experimental data, we use a next-to-leading order (NLO) QCD $K$-factor of 1.3
for any $M_{Z'}$~\cite{Gumus:2006mxa}. Apart from the QCD corrections, when various couplings of the model become large, other higher-order corrections might become important, but we have
not considered them in this simplified qualitative analysis.
\begin{figure}
\centering
\includegraphics[width=0.68\textwidth]{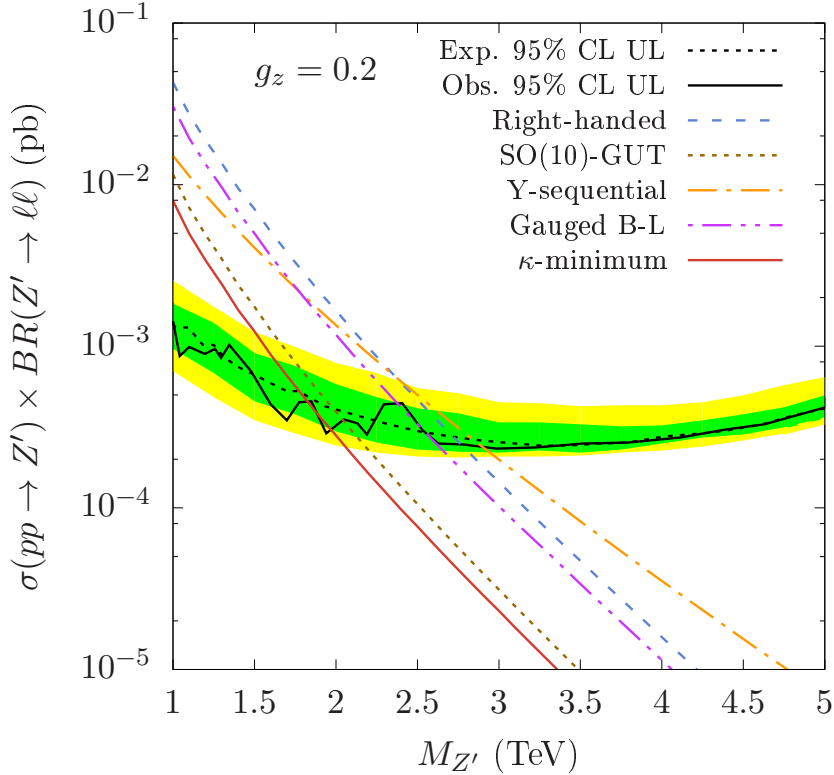}\label{fig:ATLASdata}
\caption{Comparison of the observed and expected 95\% CL UL on $\sg\times BR$ obtained from the 13 TeV ATLAS dilepton resonance search data with the theoretical predictions of various models. In this plot the reference value $g_z=0.2$ is chosen. For the Y-sequential model it is also necessary to provide a value for $z_H$; we use the reference $z_H=1$.}
\label{fig:YseqATLASCMS}
\end{figure}
In Fig.~\ref{fig:YseqATLASCMS} we compare the 95\% CL UL on the $\sg\times BR$ set by ATLAS~\cite{ATLAS:2016cyf} using dilepton data at the 13 TeV LHC with the theoretical predictions of the models discussed in section~\ref{sec:models}. We choose the benchmark value $g_z=0.2$
for this plot. Note that the dilepton BR is largely independent of $g_z$ and 
the production cross section $\sigma$ scales as $g_z^2$. Therefore it is straightforward to translate these bounds to any other choice of $g_z$.

\begin{figure}[H]
\centering
\captionsetup[subfigure]{}
\subfloat[][]{\includegraphics[width=0.4\textwidth]{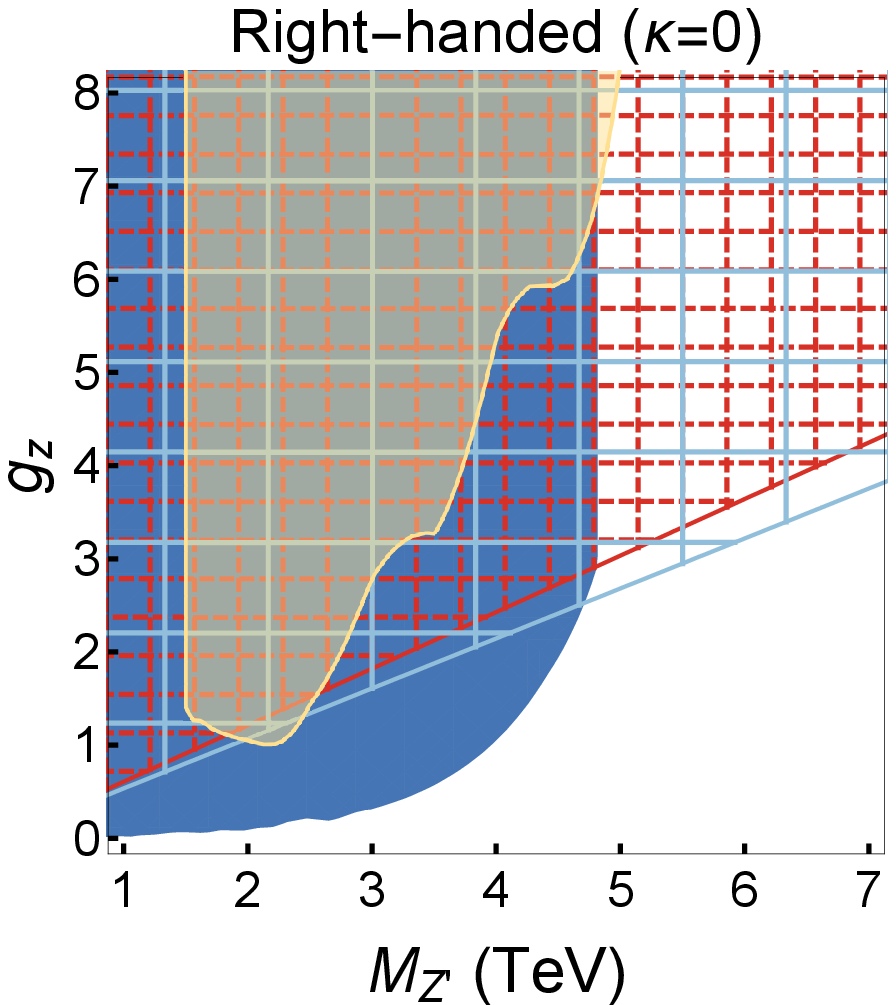}\label{sfig:R}}~~
\subfloat[][]{\includegraphics[width=0.4\textwidth]{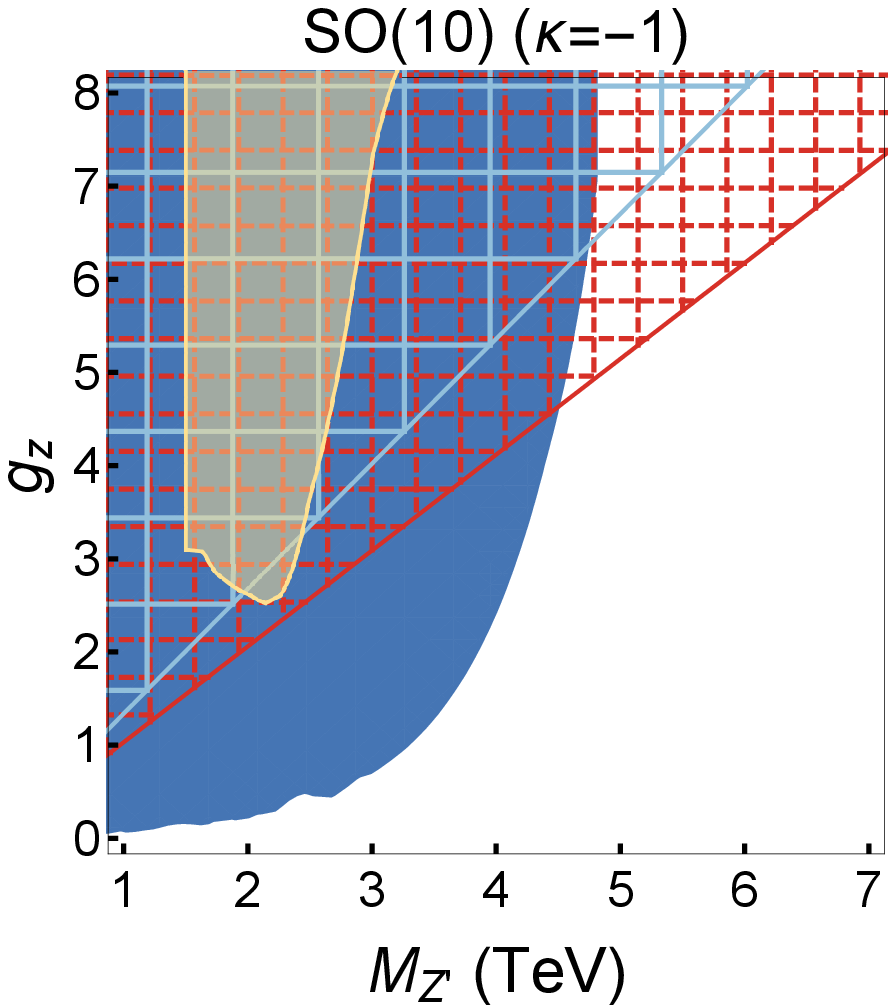}\label{sfig:SO10}}
\\
\subfloat[][]{\includegraphics[width=0.4\textwidth]{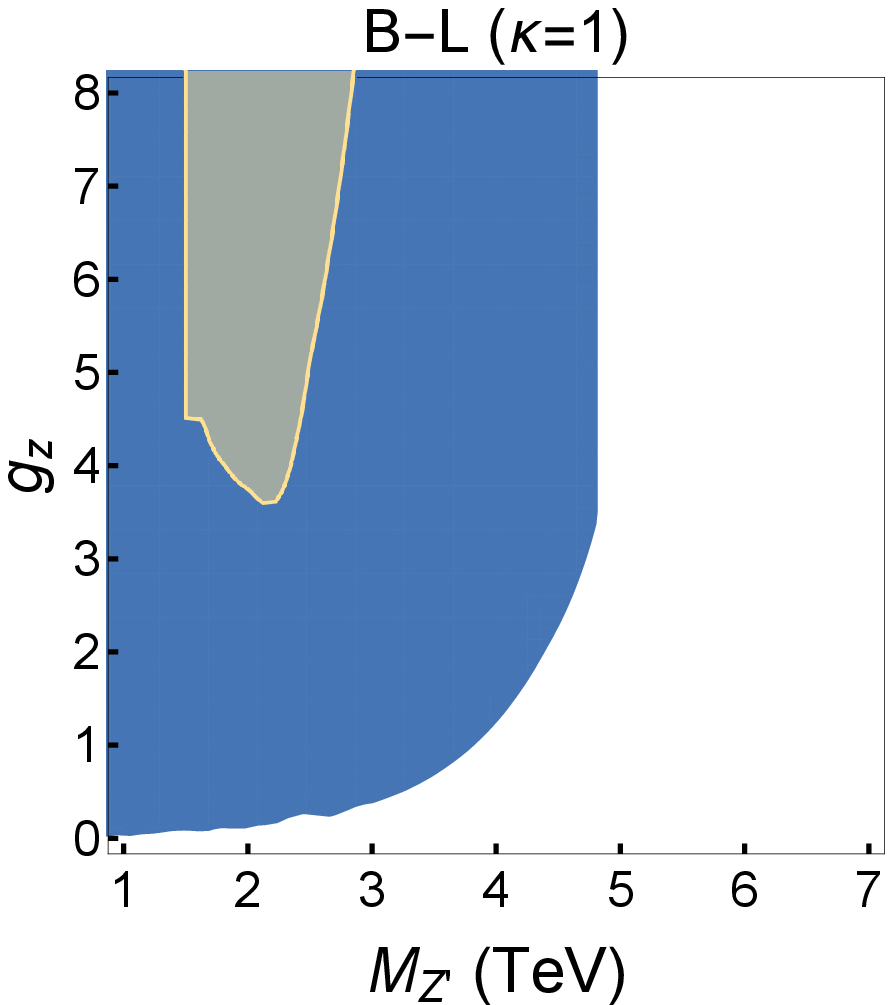}\label{sfig:BL}}~~
\subfloat[][]{\includegraphics[width=0.4\textwidth]{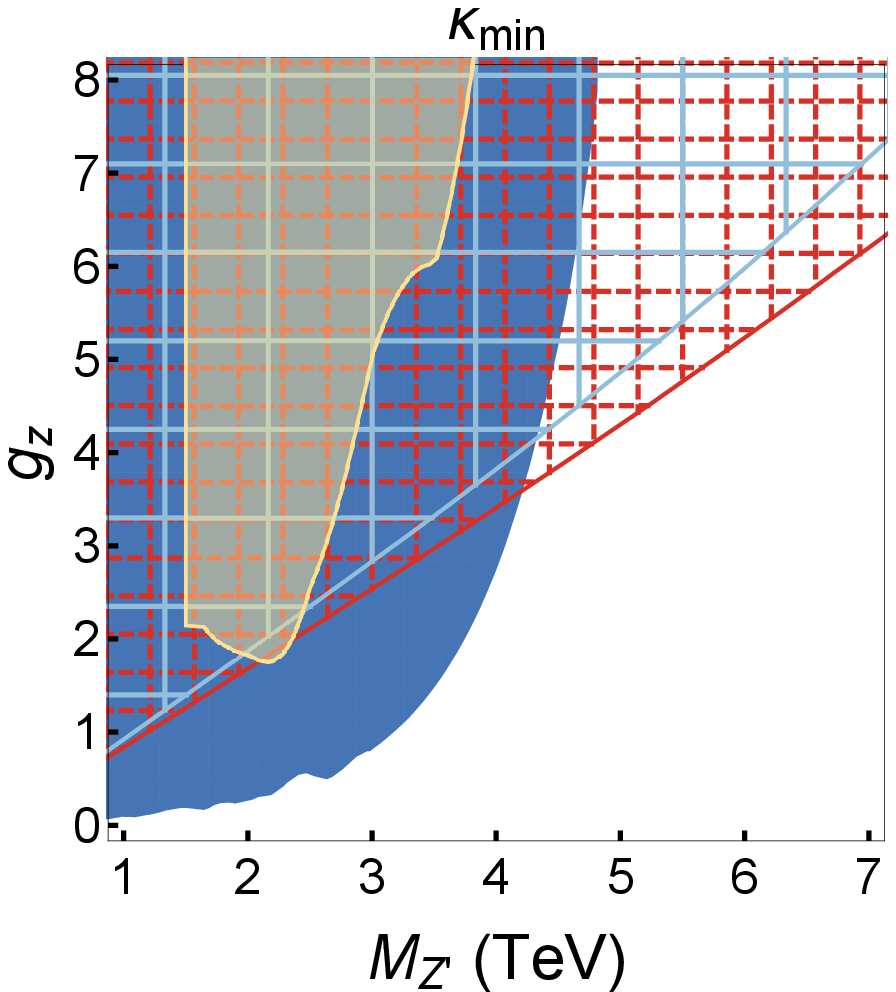}\label{sfig:kmin}}	
\label{fig:kappaExcl}
\includegraphics[width=0.8\textwidth]{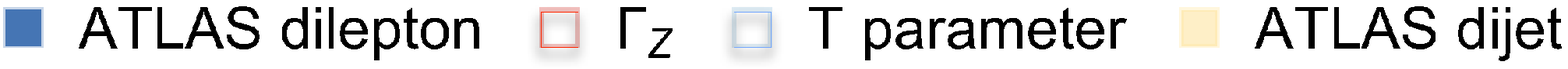}
\caption{Excluded regions. The blue filled region represents $\mc{R}>1$, where $\mc{R}=(\sg\times BR_{ll})^\text{th}/(\sg\times BR_{ll})^\text{obs}_\text{ATLAS}$; $(\sg\times BR_{ll})^\text{th}$ and 
$(\sg\times BR_{ll})^\text{obs}_\text{atlAS}$ denote the theoretical prediction and the observed
95\% CL UL set by ATLAS using dilepton data at the 13 TeV LHC, respectively. The filled beige region is the same measure but using 13 TeV ATLAS dijet data instead. The region hashed by red dashed lines corresponds to parameter points which do not fulfill the electroweak bounds set by the $T$-parameter. The region marked by light blue lines corresponds to parameter points not fulfilling the bounds set by the measured width of the $Z$-boson.
}
\label{fig:gz-MZ'}
\end{figure}

\begin{figure}[htb!]
\centering
\subfloat[][]{\includegraphics[width=0.48\textwidth]{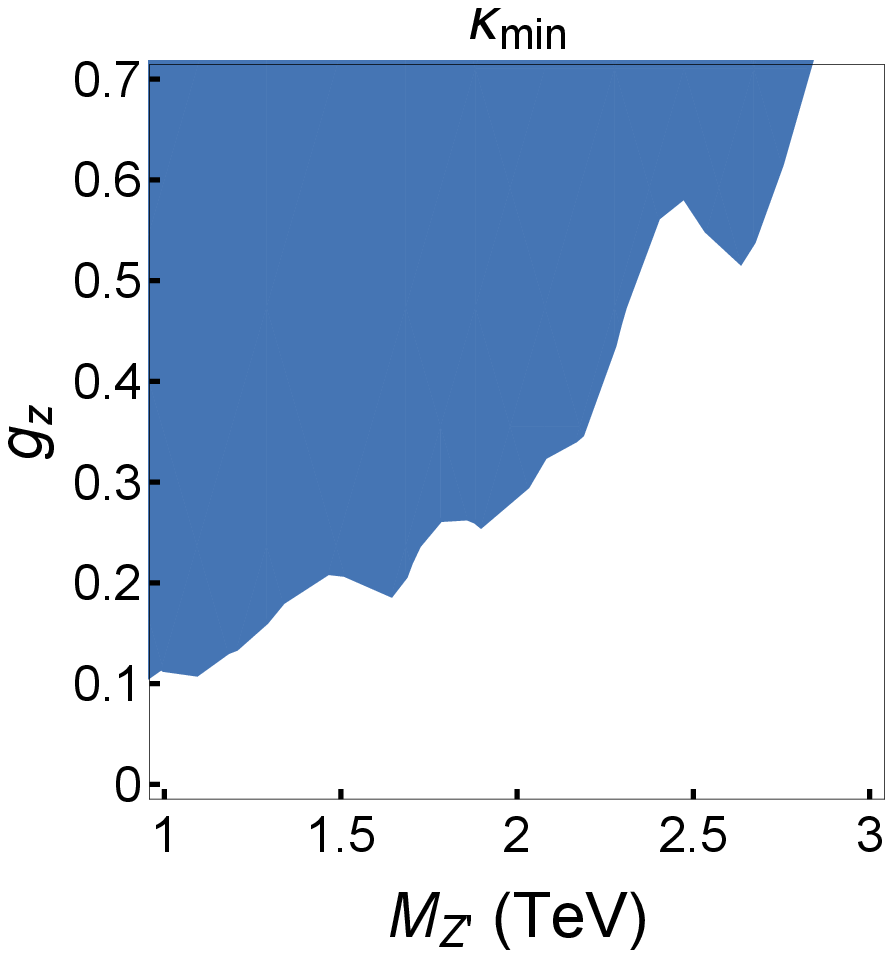}\label{fig:kminZ}}\hspace{0.25cm}
\subfloat[][]{\includegraphics[width=0.48\textwidth]{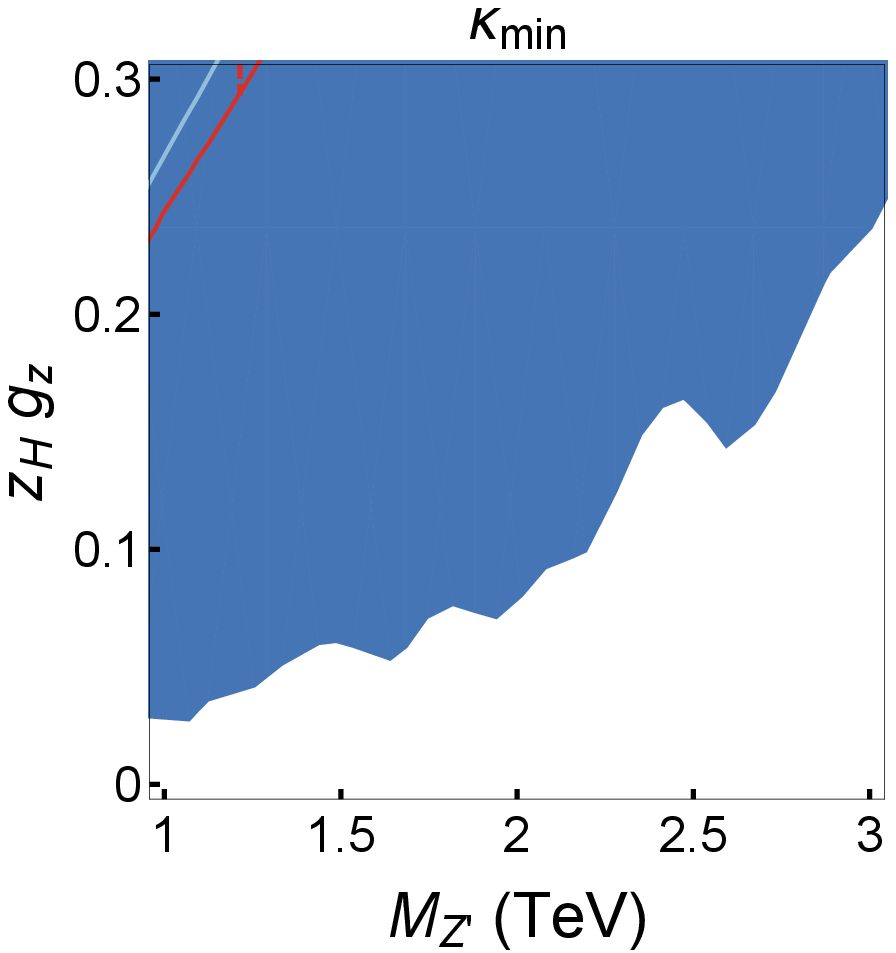}\label{fig:kminZA}}
\caption{Zoomed in version of Fig.~\ref{sfig:kmin}, with (a) $g_z$ normalization and (b) $z_H g_z$ normalization.}
\end{figure}

In Fig.~\ref{fig:gz-MZ'} we show the exclusion plots in the $g_z$ - $M_{Z'}$ plane for 
four selected models. We present exclusion regions using 13 TeV ATLAS dijet and dilepton data,
$T$-parameter constraints, and $\Gm_Z$ constraints.
The values of $\kp$ for
all the models discussed in Section~\ref{sec:models} are constant except for the 
$k_\text{min}$ model where $\kp$ varies with $M_{Z'}$; $\kp_\text{min}$ is the $\kp$-value that minimizes 
$\sg(pp\to Z')\times BR(Z'\to \ell\ell)$ for a given $M_{Z'}$. 
This implies that the excluded region for the $\kp_\text{min}$-model is also excluded for all other $\kp$ models and thus serves as a model independent upper limit of $g_z$ for a given $M_Z'$. In Fig.~\ref{fig:kminZ}, which is a zoomed in version of Fig.~\ref{sfig:kmin}, we see that for $M_{Z'} \lesssim 3 ~\text{TeV}$ the gauge coupling is constrained to $g_z \lesssim 0.8$. This bound is shown in terms of $g_z z_H$ in fig.\ref{fig:kminZA},  and the upper bound roughly correspond to $z_H g_z \lesssim 0.23$ for $M_Z' \lesssim 3~\text{TeV}$. This is a model independent upper bound on the model parameter $z_H g_z$ in this mass region.

We see from Fig.~\ref{fig:gz-MZ'} that the $g_z$ parameter space is strongly constrained from the dilepton data. Another observation is that the $B-L$ model receives no constraint from the $T$-parameter or from the $Z$ width $\Gamma_Z$, which is expected since there is no tree-level $Z-Z'$ mixing in this model. 
Note that in the $\kp$ characterization, bounds are expressed as functions of $g_z$ and $M_{Z'}$. However, the bounds on $g_z$ can be translated to bounds on $z_H g_z$ by relation 
$z_H g_z = g_z(1-\kp)/\{2(1-4\kp)\}$.

\section{Summary and conclusions}
\label{sec:discussion}

In this paper, we consider minimal anomaly free $\mathrm{U(1)}$ extensions of the SM
with a set of minimal assumptions listed in Section~\ref{sec:intro}. Apart from the 
SM particles, an electrically neutral massive $Z'$, a complex scalar $\varphi$ and
three generations of right-handed neutrinos are introduced. To make our results as model independent
as possible, we introduce a ``$\kp$-parametrization'' which is explained in Section~\ref{sec:models}. By requiring all gauge anomalies to cancel, we find that various models can be characterized by a $\kp$ value. Further requiring the model to generate Majorana masses for the right-handed neutrinos through a seesaw mechanism, the $\mathrm{U(1)}_z$ gauge charge of the Higgs  can be parametrized in terms of $\kp$. In this framework, the relevant parameters are the mass of the new gauge boson $M_{Z'}$, $\mathrm{U(1)}_z$ gauge coupling $g_z$, and the $\kp$ parameter; this parametrization is viable for all $\kp$ values except for $\kp=1/4$ (the $Y$-sequential model).
We choose the masses of the right-handed neutrinos and the new complex scalar in such a way that the decay channel is open for all considered values of $M_{Z'}$. We find that the result depends weakly on the precise values of the masses. We show that this wide class of $\mathrm{U(1)}$ extended models is mainly constrained from the new LHC dilepton data and electroweak precision measurements. 
 
The bounds on this class of models rely on the minimal assumptions outlined in Section \ref{sec:intro}. By relaxing these assumptions it could be possible to deviate from the bounds derived from data. A few possibilities are: introducing new chiral fermions that enlarge the number of possible charge assignments, allowing for generation dependent charges, considering another mechanism for EWSB, or ignoring anomaly-cancellations altogether by considering the theory as an effective field theory, perhaps supplemented by a variant of the Green-Schwarz mechanism for anomaly cancellation~\cite{Anastasopoulos:2006cz}. We will return to these issues in a forthcoming paper~\cite{inprep}.

\section*{Acknowledgments}
We thank Manuel E. Krauss and Florian Staub for helpful and interesting discussions. This work was supported by the Swedish Research Council (contract 621-2011-5107) and the Carl Trygger Foundation (contract CTS-14:206).

\end{document}